\documentclass[12pt,pra,aps,amssymb,amsfonts,amsmath,tightenlines]{revtex4}
    \usepackage{dsfont}
    \usepackage{graphicx}
    \usepackage{amssymb,amsfonts,amsthm}
    \usepackage{color}
    \usepackage{verbatim}
    \usepackage[normalem]{ulem}
    \usepackage{enumerate}
    \usepackage{mathrsfs}
    \usepackage[english]{babel}
    \usepackage[caption=false]{subfig}

    \newtheorem{Proposition}{Proposition}

    \newtheorem{Lemma}{Lemma}

    \newcommand{\re}{\mbox{$\rm e$}}
    
    \newcommand{\rd}{\mbox{$\rm d$}}

\begin{document}
    
    \title{Optimal Hedging in Incomplete Markets}
    
    \author{George Bouzianis and Lane~P.~Hughston}
    
    \affiliation{
    Department of Computing, Goldsmiths College, University of London\\ New Cross, London SE14 6NW, United Kingdom\\
    email: gbouz001@gold.ac.uk, l.hughston@gold.ac.uk 
    }
    
    \begin{abstract}
    \noindent We consider the problem of optimal hedging in an incomplete market with an established pricing kernel. In such a market, prices are uniquely determined, but perfect hedges are usually not available. We work in the rather general setting of a L\'evy-Ito market, where assets are driven jointly by an $n$-dimensional Brownian motion and an independent Poisson random measure on an $n$-dimensional state space. Given a position in need of hedging and the instruments available as hedges, we demonstrate the existence of an optimal hedge portfolio, where optimality is defined by use of an least expected squared error criterion over a specified time frame, and where the numeraire with respect to which the hedge is optimized is taken to be the benchmark process associated with the designated pricing kernel.
    \vspace{-0.2cm}
    \\
    \begin{center}
    {\scriptsize {\bf Key words: Incomplete markets, pricing kernels, hedge ratios,   Brownian motion, \\ L\'evy processes,  L\'evy measures, 
    L\'evy-Ito processes, Poisson random measure,  simulations.
    } }
    \end{center}
    \end{abstract}
    
    \maketitle
    \section {Introduction}
    \noindent This paper is concerned with optimal hedging in incomplete markets. Hedging is important, since it lies at the heart of risk management. Historically, hedging in complete markets has played a significant role in the foundations of option-pricing theory \cite{Black Scholes,  Merton 1973, Cox Ross Rubenstein, Harrison Pliska, Bensoussan}.
    From a modern perspective, however, hedging arguments need not be invoked in the determination of prices. Instead, pricing is achieved by use of a pricing kernel. The connection between the two approaches is that in a complete market the specification of the price processes of a sufficiently large number of assets is enough to allow one to determine the pricing kernel associated with that market. Nevertheless,  in the absence of market frictions, the prices of all of financial assets are determined in an incomplete market, including those of derivatives, once we designate a pricing kernel. In the incomplete market situation, however, one can not in general form a \textit{perfect} hedge of a given position. This leaves us with a more precise statement of our problem: namely, determination of the \textit{optimal} strategy for hedging a financial position in an incomplete market, given the set of hedging assets at the hedger's disposal. The optimal hedge corresponds to the maximal possible elimination of risk in a financial position making use of the instruments available for this purpose.
    
    The paper is structured as follows. In Section II we briefly summarize some of the mathematical ideas that we require. We define what we mean by a L\'evy-Ito process and in Proposition 1 we recall the general form of Ito's formula applicable to L\'evy-Ito processes.  Then in Proposition 2 we give a version of the Ito formula that holds when the large jumps are moderated, which is useful in financial applications. In Proposition 3 we comment on the form that the Ito isometry takes in the L\'evy-Ito setting. 
    In Section III we introduce the family of risky assets that we work with in the hedging problem. We argue that the most natural approach to hedging arises when the values of the various assets under consideration are expressed in units of the benchmark process associated with the pricing kernel. In Section IV we consider the hedging of a position in a risky asset in a one-dimensional L\'evy-Ito market in the situation where the hedging instrument is another risky asset driven by the same one-dimensional 
    L\'evy-Ito process. In general, a perfect hedge is not possible in such a market, so one aims for a best possible hedge instead. We take the view that the goal is that of optimal elimination of the risk, which we characterize in a natural way using a quadratic optimization criterion.  See \cite{Arai 2005, Biagini Oksendal 2006, Cerny Kallsen 1984, Cont Tankov Voltchkova, Delbaen Schachermayer 1996, Follmer Sondermann 1986,
    Gourieroux Laurent Pham 1998, Hubalek Kallsen Krawczyk 2006, Lim 2006, Pham 2000, 
    Schweizer 2001} for aspects of quadratic hedging. In Proposition 4 we obtain a formula for the optimal hedge in the case of a single hedging asset. We refer to the asset being hedged as the contract asset. The terminology is inherited from the language of derivatives pricing, though in the present context the asset being hedged need not be a derivative; indeed, the various assets involved are essentially on an equal footing.  In Section V we consider the situation where we hedge the contract asset with a position in $n$ risky assets. In Proposition 5 we work out an expression for the optimal hedge in such a market, and in Proposition 6 we show that if there is negligible redundancy among the hedging assets then the optimal hedge obtained with $n+1$ hedging instruments is better than the optimal hedge obtained with $n$ such instruments. 
In Section VI we look in more detail at the case where two hedging assets are available to hedge the contract asset, and an explicit formula for the optimal hedge is given in Proposition 7. We illustrate the results in the simplest possible situation: this is the case of a geometric L\'evy asset for which  the L\'evy process is a linear combination of a Brownian motion and a Bernoulli process.  We refer to a L\'evy process of this type as a Bernoulli jump diffusion. By a Bernoulli process we mean a compound Poisson process for which each jump is characterized by an independent Bernoulli random variable taking one of two possible values.  We consider the situation where  the contract asset and the hedging assets are geometric Bernoulli jump diffusions driven by the same L\'evy-Ito process. We illustrate the fundamental fact that a better hedge can be obtained by using both of the hedging assets rather than just a single hedging asset, even though a perfect hedge is not obtainable as long as the Brownian component of the driving process is present. On the other hand, if the Brownian volatility is small for the various assets under consideration, then a nearly perfect hedge can be obtained. 
        Finally, we set out some useful formulae from the L\'evy-Ito calculus in an Appendix. 
    
    \section{Mathematical preliminaries}
    
    \noindent We begin with a brief account of the mathematical context in which we formulate the hedging problem. Most of the material in this section is well known, but we find it convenient to set out various details. The L\'evy-Ito market provides a modelling framework of considerable generality. In particular, it contains all of the familiar Brownian motion driven models and L\'evy driven models as special cases. The setup is as follows. We fix a probability space $(\Omega, \mathscr{F} , \mathbb{P})$ that supports an $n$-dimensional Brownian motion $\{W_t\}_{t\geq0}$ alongside an independent  Poisson random measure $\{{N}({\rm d}x, {\rm d}t)\}$ with mean measure  $\nu({\rm d}x)\, {\rm d}t$, where $\nu({\rm d}x)$ is taken to be the L\'evy measure associated with an $n$-dimensional pure-jump L\'evy process. Thus $\nu({\rm d}x)$ is a $\sigma$-finite measure on $(\mathbb{R}^n, \mathscr{B}(\mathbb{R}^n))$ such that $\nu(\{0\}) = 0 $ and 
    \begin{eqnarray}
    \int_{\mathbb{R}^n} \min\left(1, |x|^2 \right)  \nu({\rm d}x) < \infty \,.
    \end{eqnarray}
    We write $\{\mathcal F_t\}_{t\geq0}$ for the augmented filtration generated by $\{W_t\}$ and $\{{N}({\rm d}x, {\rm d}t)\}$. See \cite{Applebaum, Oksendal Sulem 2014, Jeanblanc Yor Chesney, BHJS,  Eberlein Kallsen}
     for aspects of the theory of L\'evy-Ito processes. In the one-dimensional case, by a L\'evy-Ito process driven by $\{W_t\}$ and $\{{N}({\rm d}x, {\rm d}t)\}$ we mean a process $\{X_t\}_{t\geq0}$ satisfying a dynamical equation of the form
    \begin{eqnarray} 
    {\rm d}X_t =  \alpha_t \, {\rm d}t +  \beta_t \, {\rm d}W_t + \int_{|x|\in(0,1)} \gamma_t(x) \, \tilde{{N}}({\rm d}x, {\rm d}t) + \int_{|x|\geq1} \delta_t(x) \, {N}({\rm d}x, {\rm d}t)\,,
    \label{Levy Ito process differential form}
    \end{eqnarray}
    where
    \begin{eqnarray} 
    \tilde{ N}({\rm d}x, {\rm d}t) = {N}({\rm d}x, {\rm d}t) - \nu({\rm d}x)\,{\rm d}t\,.
    \label{compensated PRM}
    \end{eqnarray}
    We require that $\{ \alpha_t \}_{t \geq 0}$ and $\{ \beta_t \}_{t \geq 0}$ be $\{\mathcal F_t\}$-adapted, that $\{ \gamma_t(x)\}_{t \geq 0,\,|x| <1}$ and $\{ \delta_t(x)\}_{t \geq 0,\,|x| \geq 1}$  be $\{\mathcal F_t\}$-predictable, and that  
    \begin{eqnarray} 
     \mathbb{P} \left[  \,
    \int_{0}^{t} \left(  |\alpha_s| +  \beta_s^{\,2} +
    \int_{|x|<1} \gamma_s(x)^{\,2} \, \nu({\rm d}x)  \right) \, {\rm d}s < \infty \right ] = 1
     \label{P condition}
    \end{eqnarray}
    for $t \geq 0$.  We note that the integral with respect to $\tilde{ N}({\rm d}x, {\rm d}t)$ in equation \eqref{Levy Ito process differential form} and similar expressions of this type is defined by means of a limiting procedure as the origin is approached, as described, e.g., in reference \cite{Sato 1990} at page 120. Then  we have the following generalization of Ito's formula (see, for example, reference \cite{Applebaum}, Theorem 4.4.7):
    \vspace{0.2 cm}
    \begin{Proposition} \label{Ito formula} Let $F: \mathbb{R} \rightarrow \mathbb{R}$ admit a continuous second derivative and let $\{X_t\}$ be a L\'evy-Ito process for which the dynamics are as in \eqref{Levy Ito process differential form}. Then for $t \geq 0$ it holds that
    \begin{align} 
    {\rm d} F(X_t) & =  \left[ \alpha_t \, F{'}(X_{t^{-}}) + \frac{1}{2} \beta_t^{\,2} \, F{''}(X_{t^{-}})\right] {\rm d}t + \beta_t \,F{'}(X_{t^{-}}) \,{\rm d}W_t \nonumber
    \\ & \hspace{1cm} + \int_{|x|<1}\left[F(X_{t^{-}} + \gamma_t(x)) - F(X_{t^{-}}) - \gamma_t(x)F{'}(X_{t^{-}})\right]\, \nu({\rm d}x) {\rm d}t \nonumber
    \\ &\hspace{1cm} + \int_{|x|\in(0,1)}\left[F(X_{t^{-}} + \gamma_t(x)) - F(X_{t^{-}})\right]\, \tilde{N}({\rm d}x, {\rm d}t ) \nonumber
    \\ & \hspace{1cm} + \int_{|x| \geq 1}\left[F(X_{t^{-}} + \delta_t(x)) - F(X_{t^{-}})\right]\, N({\rm d}x, {\rm d}t ) \, .
    \end{align}
    \end{Proposition}
    \vspace{0.2cm}
    \noindent We can use the generalized Ito formula to work out Ito product and quotient rules for such processes. These results are very useful, so for the convenience of the reader we set them down in detail in the Appendix. 
    In some situations will be appropriate to consider processes for which the dynamical equation takes the form 
    \begin{eqnarray} 
    {\rm d}X_t =  \alpha_t \, {\rm d}t +  \beta_t \, {\rm d}W_t + \int_{|x|\in(0,1)} \gamma_t(x) \, \tilde{{N}}({\rm d}x, {\rm d}t) + \int_{|x|\geq1} \delta_t(x) \, \tilde{N}({\rm d}x, {\rm d}t)\,,
    \label{Levy Ito process symmetrical}
    \end{eqnarray}
    where the integral involving the large jumps is taken with respect to the compensated Poisson random measure. In order for this to be possible, $\{\delta_t(x)\}$ must satisfy 
    \begin{eqnarray} 
    \mathbb P \left[ \int_{|x|\geq1} | \delta_t(x) | \, {\nu}({\rm d}x) < \infty\right] = 1\,,
    \label{condition on delta}
    \end{eqnarray}
    which is sufficient to ensure that the integral with respect to the compensated Poisson random measure exists for large jumps. If we impose the stronger condition 
    \begin{eqnarray} 
    \mathbb P \left[ \int_{|x|\geq1} \delta_t(x)^{\,2} \, {\nu}({\rm d}x) < \infty\right] = 1\,,
    \label{quadratic condition on delta}
    \end{eqnarray}
    we can simplify and unify the notation by using a common symbol $\{\gamma_t(x)\}_{t \geq 0, \, x \in \mathbb{R}}$ for the coefficients of the compensated Poisson random measures for small jumps and large jumps. Then we write
    \begin{eqnarray} 
    {\rm d}X_t =  \alpha_t \, {\rm d}t +  \beta_t \, {\rm d}W_t + \int_{|x|>0} \gamma_t(x) \, \tilde{{N}}({\rm d}x, {\rm d}t) \,,
    \label{Levy Ito process completely symmetrical}
    \end{eqnarray}
    and the associated condition on the coefficients takes the form
    \begin{eqnarray} 
     \mathbb{P} \left[  \,
    \int_{0}^{t} \left(  |\alpha_s| +  \beta_s^{\,2} +
     \int_{x} \gamma_s(x)^{\,2} \, \nu({\rm d}x)  \right) \, {\rm d}s < \infty \right ] = 1\,,
     \label{P condition symmetrical}
    \end{eqnarray}
    in place of \eqref{P condition}, where the subscript $x$ denotes integration over the whole of the real line.
    We shall refer to processes satisfying \eqref{Levy Ito process completely symmetrical} and \eqref{P condition symmetrical} as being ``symmetric" since large and small jumps are treated similarly. Symmetric processes turn out to be useful in financial applications, where the stronger condition on the integrability of the jump volatility with respect to the L\'evy measure for large jumps is not unreasonable. In the symmetric case Ito's formula takes the following form:
    \vspace{0.2cm}
    \begin{Proposition} \label{Symmetric Ito formula} Let $F: \mathbb{R} \rightarrow \mathbb{R}$ admit a continuous second derivative and let $\{X_t\}$ be a symmetric L\'evy-Ito process for which the dynamics are as in  \eqref{Levy Ito process completely symmetrical}. Then for $t \geq 0$ we have
    \begin{align} 
    {\rm d} F(X_t) & =  \left[ \alpha_t \, F{'}(X_{t^{-}}) + \frac{1}{2} \beta_t^{\,2} \, F{''}(X_{t^{-}})\right] {\rm d}t + \beta_t \,F{'}(X_{t^{-}}) \,{\rm d}W_t \nonumber
    \\ & \hspace{1cm} + \int_{x}\left[F(X_{t^{-}} + \gamma_t(x)) - F(X_{t^{-}}) -  \gamma_t(x)F{'}(X_{t^{-}})\right]\, \nu({\rm d}x) {\rm d}t \nonumber
    \\ &\hspace{1cm} + \int_{|x|>0}\left[F(X_{t^{-}} + \gamma_t(x)) - F(X_{t^{-}})\right]\, \tilde{N}({\rm d}x, {\rm d}t ).
    \end{align}
    \end{Proposition}
    \noindent The higher-dimensional analogues of Propositions 1 and 2 are straightforward. Finally, we note that the Ito isometry can be generalized in  the present context. So far, we have not imposed any integrability conditions on the processes that we have considered. For the L\'evy-Ito analogue of the Ito isometry we require that the process satisfies an $L^2$ condition. 
    \vspace{0.1 cm}
    \begin{Proposition} \label{Proposition Ito isometry} Let $\{ X_t\}_{t\geq 0}$ be a L\'evy-Ito process such that
    \begin{eqnarray} 
    X_t = X_0 + \int_0^t   \beta_s \, {\rm d}W_s + \int_0^t  \int_{|x|>0} \gamma_s(x) \, \tilde{{N}}({\rm d}x, {\rm d}s) \, ,
    \end{eqnarray}
    where $X_0$ is a constant and
    \begin{eqnarray} 
     \mathbb{P} \left[  \,
    \int_{0}^{t} \left(   \beta_s^{\,2} +
     \int_{x} \gamma_s(x)^{\,2} \, \nu({\rm d}x)  \right) \, {\rm d}s < \infty \right ] = 1 \,.
    \end{eqnarray}
    If \,
    $\mathbb{E}\left[X_t^{\,2}\right] < \infty$ for $t\geq0$, then $\{ X_t\}_{t\geq 0}$ is a martingale and for $t\geq0$ it holds that
    \begin{eqnarray} 
    \mathbb{E}\left[(X_t - X_0)^2\right] =
     \mathbb{E} \left[  \,
    \int_{0}^{t} \left(   \beta_s^{\,2} +
     \int_{x} \gamma_s(x)^{\,2} \, \nu({\rm d}x)  \right)  {\rm d}s \right ] .
    \label{Ito isometry}
    \end{eqnarray}
    \end{Proposition}
    \vspace{0.1 cm}
    \noindent Again, the corresponding result for an $n$-dimensional L\'evy-Ito process is straightforward.
    \section{Risky assets}
    \noindent We proceed to consider the problem of optimal hedging. It should be emphasized from the outset that we are not concerned here with the problem of derivative pricing via hedging arguments. We assume that prices are known and we look instead at the problem of hedging a position in one asset by use of a self-financing portfolio of other assets. In a complete market we know that an exact hedge can be obtained in such a situation; but we work in an incomplete market, where exact hedges are generally not available, so we look for an optimal hedge instead.  
    We fix a probability space $(\Omega, \mathcal{F}, \mathbb{P})$ where $\mathbb{P}$ is the real-world measure. The market filtration $\{\mathcal F_t\}_{t \geq 0}$ is taken to be the augmented filtration generated by a one-dimensional Brownian motion $\{W_t\}$ and an independent one-dimensional Poisson random measure $\{{N}({\rm d}x, {\rm d}t)\}$, where the Poisson random measure is that associated with a one-dimensional pure-jump L\'evy process in the sense discussed in Section II.  
    
    We introduce a fiat currency, which we call the domestic currency, in units of which prices are conventionally expressed. The market is assumed to be endowed with a pricing kernel $\{\pi_t\}_{t\geq 0}$ for which the dynamics take the form
    \begin{eqnarray}
    {\rm d} \pi_t  = - \pi_{t^{-}} \left[ r_t \, {\rm d}t + \lambda_t \,  {\rm d}W_t + \int_{|x|>0} \Lambda_t(x) \, \tilde{N}({\rm d}x, {\rm d}t)  \right ] .
    \label{pricing kernel dynamics}
    \end{eqnarray} 
    We assume that the domestic short rate $\{r_t\}_{t \geq 0}$ and the Brownian market price of risk $\{\lambda_t\}_{t \geq 0}$ are adapted and that the jump market price of risk $\{\Lambda_t(x)\}_{t \geq 0,\, x \in \mathbb R}$ is predictable and such that 
    $\Lambda_t(x) < 1$ for $t\geq 0$ and $x \in \mathbb R$ . The solution for the pricing kernel is then 
    \begin{align}
    \pi_t = & \exp \bigg[-\int_0^t r_s \, {\rm d}s  - \int_{0}^{t} \lambda_s \,  {\rm d}W_s - \frac{1}{2} \int_{0}^{t} \lambda_s^{\,2}  \, {\rm d}s
     \nonumber \\&\quad \quad - \int_{0}^{t} \int_{|x|>0} \kappa_s (x) \, \tilde{N}({\rm d}x, {\rm d}s) - \int_{0}^{t} \int_{x} \left(\re^{-\kappa_s(x)} -1 + \kappa_s(x) \right) \, \nu({\rm d}x) \, {\rm d}s  \bigg] ,
    \label{Levy Ito pricing kernel RA}
    \end{align} 
    where $\{\kappa_t(x)\}_{t \geq 0,\, x \in \mathbb R}$ is defined by 
    \begin{align}
    \kappa_t(x) = \log \left[ \frac {1}{1 - \Lambda_t(x)}\right] .
    \end{align} 
    We assume that the market includes a money market asset $\{B_t\}_{t \geq 0}$ satisfying ${\rm d}B_t = r_t \, B_t\,{\rm d}t $,
    along with one or more risky assets. For a typical risky asset we let $\{S_t\}_{t \geq 0}$ denote the price process, and for simplicity we assume that the asset pays no dividend. The associated dynamics are taken to be of the form
    \begin{align} 
    \frac{{\rm d}S_t}{\,\,\,S_{t^{-}}} = \bigg[ r_t   + \lambda_t \, \sigma_t  + \int_{x} \Lambda_t (x) \, \Sigma_t (x)\, \nu({\rm d}x)\, \bigg]  {\rm d}t  +  \sigma_t \,  {\rm d}W_t + \int_{|x|>0} \Sigma_t(x) \, \tilde{{N}}({\rm d}x, {\rm d}t)\, ,
    \label{Asset dynamics}
    \end{align}
    where $\{\sigma_t\}_{t \geq 0}$ is adapted, $\{\Sigma_t(x)\}_{t \geq 0,\, x \in \mathbb R}$ is predictable, and  $\Sigma_t(x) > -1$ for $t\geq 0$ and $x \in \mathbb R$. We shall require that the dynamics of $\{S_t\}$ are non-degenerate in the following sense. Let $\mathscr D$ denote the subset of $\Omega \times [0, T]$  over which it holds that  
 \begin{eqnarray}  
    \sigma_t ^{\,2 }+ \int_{x} \Sigma_t(x) ^2 \, \nu({\rm d}x)=0 \,.
    \label{degeneracy, one asset}
 \end{eqnarray} 
 We say that $\{S_t\}$ has non-degenerate dynamics if $\mathscr D$ has $\{\mathbb P \times {\rm Leb}\,[0,T]\}$ measure 0. An alternative way of expressing the non-degeneracy condition is as follows. Let the support of the L\'evy measure $\nu(\rd x)$ be defined as the set $S_\nu$ comprising all $x \in \mathbb R$ such that $\nu(A) > 0$ for any open set $A$ containing $x$ (\cite{Sato 1990}, page 148). Then $\mathscr D \subset \{\Omega \times [0, T]\}$ can be defined to be the set 
  \begin{eqnarray}
 \mathscr D = \{ \omega \in \Omega, t \in [0, T] :   \sigma_t = 0 \, \cap \, \Sigma_t(x) = 0, \, x \in S_\nu \}.
  \end{eqnarray}
It should be evident that these definitions of the degeneracy subset are equivalent, and it is useful to keep both in mind. 

We require that for any risky asset the process determined by the product of the pricing kernel and the asset price should be a $\mathbb P$-martingale. Thus we have
    \begin{eqnarray}
    S_t = \frac{1}{\pi_t}\,\mathbb{E}_t [\pi_u S_u]
    \end{eqnarray}
    for $0\leq t \leq u < \infty$, where $\mathbb{E}_t$ denotes conditional expectation with respect to $\mathcal F_t$. 
    There is another way of expressing this condition which turns out to be useful for our purposes. It is well known that the process $\{\xi_t\}_{t \geq 0}$ defined by $\xi_t = 1/ \pi_t$ for $t \geq 0$ can be interpreted as a ``natural numeraire" or ``benchmark". By the definition of the pricing kernel, we see that for any asset $\{S_t\}$ that pays no dividend the process $\{\bar S_t\}$ defined by $\bar S_t = S_t / \xi_t$ represents the price of the original asset expressed in units of the natural numeraire. It follows that the ``natural" price of any such asset is a martingale. Then we have
    \begin{eqnarray}
    S_t = \xi_t \, \mathbb{E}_t \left[ \frac{S_u}{\xi_u} \right]
    \label{fluctuating term}
    \end{eqnarray}
    for $0\leq t \leq u < \infty$, or equivalently 
    \begin{eqnarray}
    \bar S_t = \mathbb{E}_t \left[\bar S_u\right].
    \label{S bar}
    \end{eqnarray}
    Equation \eqref{fluctuating term} shows that the domestic value of the asset at time $t$ can be represented as the product of the natural numeraire (which can be interpreted as a dividend-adjusted proxy for the market as a whole) and a fluctuating term, given by the conditional expectation of the natural value of the asset at some later time $u$. 
    A form of  \eqref{S bar} is used in the theory of derivatives, for instance, when we make use of the pricing formula
    \begin{eqnarray}
     H_t = \frac{1}{\, \pi_t \,} \, \mathbb{E}_t \left[ \pi_T H_T\right],
    \end{eqnarray}
    valid for $0 \leq t < T < \infty$, which shows that the natural value $\bar H_t = \pi_t H_t$ of the derivative at $t$ is given by the conditional expectation of the natural value of the payoff $\bar H_T = \pi_T H_T$.  
    
    A calculation making use of \eqref{pricing kernel dynamics}, \eqref{Asset dynamics} and Lemma \ref{proportional symmetric product rule} (see Appendix) shows that the dynamical equation satisfied by the natural value of the risky asset takes the form 
    \begin{eqnarray}
    \frac{{\rm d}\bar S_t}{\,\,\,\bar S_{t^{-}}} =  \bar \sigma_t \,  {\rm d}W_t + \int_{|x|>0} \bar \Sigma_t(x) \, \tilde{{N}}({\rm d}x, {\rm d}t)\, ,
    \label{Natural Asset price}
    \end{eqnarray} 
    where $\bar \sigma_t = \sigma_t - \lambda_t$ and  
    $\bar \Sigma_t(x) =  \Sigma_t(x)\big( 1 - \Lambda_t(x) \big) - \Lambda_t(x)$,
    or equivalently
    \begin{align} 
    \sigma_t = \bar \sigma_t + \lambda_t, \quad  \Sigma_t(x) = 
    \frac {\bar \Sigma_t(x) + \Lambda_t(x)} {1 - \Lambda_t(x)} \,.
    \label{decomposition}
    \end{align}
    The relations given in \eqref{decomposition} demonstrate that the  Brownian and jump volatilities of the asset with domestic price process 
    $\{S_t\}$ can each be decomposed into terms involving only the intrinsic ``natural" volatility of the asset and terms associated with the volatility of the domestic pricing kernel but not associated with any particular asset.
    
    One can check that as a consequence of \eqref{pricing kernel dynamics} and Proposition \ref{Symmetric Ito formula} (or Lemma 6), the dynamical equation satisfied by $\{\xi_t\}$ takes the form
    \begin{align} 
    \frac{{\rm d}\xi_t}{\,\,\,\xi_{t^{-}}} =  \, \bigg[ r_t \,  + \lambda_t^{\,2} \, + \int_{x} \frac{\Lambda_t (x) ^2}{1-\Lambda_t (x)} \, \nu({\rm d}x)\, \bigg]  {\rm d}t  + \, \lambda_t \,  {\rm d}W_t + \int_{|x|>0} \frac{\Lambda_t (x)}{1-\Lambda_t (x)} \, \tilde{{N}}({\rm d}x, {\rm d}t)\, ,
    \label{natural numeraire dynamics}
    \end{align}
    which is indeed of the type appropriate to an asset that pays no dividend, as one sees by comparing \eqref{Asset dynamics} with \eqref{natural numeraire dynamics}. The benchmark process has the property that its Brownian proportional volatility coincides with the Brownian market price of risk and its jump proportional volatility is given by an invertible function of the jump market price of risk. 
    
    The significance of the benchmark asset in the present investigation is as follows. We are concerned with the problem of hedging a position in a risky asset with a position in a portfolio consisting of one or more other risky assets. Now, when such a hedge is carried out, this involves a choice of base currency with respect to which the hedge is optimized. Clearly, the choice of base currency is largely arbitrary, and it does not make sense to insist on minimizing exclusively the magnitude of the residual value of the hedge portfolio in units of the domestic currency. Sometimes it is argued that there may be a favoured choice of base currency -- for example the currency in which a household has to meet its daily obligations, or in which a business has to accommodate a series of cashflows in connection with its activities. But such considerations bring additional elements of structure  into the argument, and the fact remains that there is no \textit{a priori} reason why one fiat currency should be favoured over another in the absence of a more detailed specification of the problem. Of all the choices of hedging currencies there is, however, a ``preferred" numeraire involving no additional elements of structure, and this is the benchmark. So we take the view that the optimization problem takes the form of minimizing a function of the magnitude of the value of the hedge portfolio when that value is expressed in units of the benchmark.

    Proceeding with our investigation of optimal hedging, let us write $\{C_t\}_{t \geq 0}$ for the domestic price process of another risky asset, which we call the {\em contract asset}. We shall assume that $\{C_t\}$ is strictly positive and that 
    \begin{align} 
    \frac{{\rm d}C_t}{\,\,\,C_{t^{-}}} = \bigg[ r_t   + \lambda_t \, \sigma_t^c   + \int_{x} \Lambda_t (x) \, \Sigma_t^c (x)\, \nu({\rm d}x)\, \bigg]  {\rm d}t  +  \sigma_t^c \,  {\rm d}W_t + \int_{|x|>0} \Sigma_t^c(x) \, \tilde{{N}}({\rm d}x, {\rm d}t)\, ,
    \label{Asset to hedge dynamics}
    \end{align}
    where $\{\sigma^c_t\}_{t \geq 0}$ is adapted, $\{\Sigma^c_t(x)\}_{t \geq 0,\, x \in \mathbb R}$ is predictable,  and $\Sigma^c_t(x) > -1$  for $t\geq 0$ and $x \in \mathbb R$.
    We can think of $\{C_t\}$ as representing the domestic value process of the position that we wish to hedge, and $\{S_t\}$ as being the domestic value process of the hedging asset.

    For applications, one usually needs to impose stronger conditions on the price processes under consideration. For example, in the case of a derivative, with payoff $H_T$ at time $T$, it is reasonable to assume not merely that the payoff should satisfy $\mathbb{E}\,[\bar H_T] < \infty$, but also that it should satisfy $\mathbb{E}\,[\bar H_T^{\,2}] < \infty$. In other words, for derivative risk management, we typically desire that some measure of the uncertainty of the payoff can be worked out, such as its variance. Indeed, in financial markets, one does not really wish to be working with instruments that are so volatile or ill-behaved that it is not possible to assign a meaningful value to the variance of the payoff. Since, in international markets, there is no particular reason to prefer one currency to another, it makes sense to introduce a minimalist assumption to the effect that the variance of the natural value of the payoff should be quantifiable. Thus, we shall assume at the very least that ${\rm Var} \, \bar H_T < \infty$. One could consider other choices for a measure of the riskiness of the payoff, and one could work this out in other units, but the choice that we have indicated is convenient from a mathematical perspective since the category of square-integrable random variables is well understood, and the use of natural units is well defined already under the assumptions that we have made. One might object that insisting on a finite variance is too strong an assumption; but the reply can be put in normative terms -- namely, that for a financial instrument to be considered as a legitimate object of commerce, it needs in principle to be capable of being risk-managed in a reasonably conventional manner; and the requirement that the value of the  instrument can be modelled as having a finite variance is a step in this direction, an embodiment of this idea. 
    \section{Optimal Hedging in a L\'evy-Ito market}
    %
    \noindent We consider setting up a trading strategy to hedge the natural value of a position in a given asset. Going forward, we shall for this purpose assume that all values are given in natural units -- that is, in units of the natural benchmark numeraire. Thus, we henceforth drop the use of the ``bar'' notation, and let $\{S_t\}$ and $\{C_t\}$ denote the \emph{natural} prices of the hedging asset and the contract asset, respectively. For the associated price dynamics we write 
    \begin{eqnarray}
    \frac{{\rm d} S_t}{\,\,\, S_{t^{-}}} =  \sigma_t \,  {\rm d}W_t + \int_{|x|>0} \Sigma_t(x) \, \tilde{{N}}({\rm d}x, {\rm d}t)\,
    \label{Natural Asset price bar drop}
    \end{eqnarray} 
    and
    \begin{eqnarray}
    \frac{{\rm d} C_t}{\,\,\, C_{t^{-}}} = \sigma^{c}_t \,  {\rm d}W_t + \int_{|x|>0}  \Sigma^{c}
    _t(x) \, \tilde{{N}}({\rm d}x, {\rm d}t)\,.
    \label{Second Natural Asset price bar drop}
    \end{eqnarray} 
    Writing 
    \begin{eqnarray}
    \sigma_t (x) = \log \left (1 + \Sigma_t(x)\right ), \quad \sigma^c_t (x) = \log \left (1 + \Sigma^c_t(x)\right ), 
    \end{eqnarray} 
    one can use the Proposition 2 to show that the corresponding price processes are given by the expressions
    \begin{align}
    	S_t &= S_0 \exp\left(\int_{0}^{t} \sigma_u \,{\rm d}W_u - \frac{1}{2}\int_{0}^{t} \sigma^{\,2}_u\, {\rm d}u \right) \nonumber
    	\\
    	& \hspace{0.6cm}
    	\times \exp\left(\int_{0}^{t}\int_{|x|>0} \sigma_u(x) \tilde{N}({\rm d}x, {\rm d}u) -\int_{0}^{t}\int_{x} (\re^{\sigma_u(x)} -\sigma_u(x) -1)\nu({\rm d}x){\rm d}u \right)
    	\label{Natural Asset price process bar drop}
    \end{align} 
    and
    \begin{align}
    	C_t &= C_0 \exp\left(\int_{0}^{t} \sigma^c_u \,{\rm d}W_u - \frac{1}{2}\int_{0}^{t} \,(\sigma^c_u)^2 {\rm d}u \right) \nonumber
    	\\
    	& \hspace{0.6cm}
    	\times \exp\left(\int_{0}^{t}\int_{|x|>0} \sigma^c_u(x) \tilde{N}({\rm d}x, {\rm d}u) -\int_{0}^{t}\int_{x} (\re^{\sigma^c_u(x)} -\sigma^c_u(x) -1)\nu({\rm d}x){\rm d}u \right).
    	\label{Second Natural Asset price process bar drop}
    \end{align} 
    The hedging problem can be formulated as follows. The hedger is holding a position in one unit of the contract asset. The value process of this asset is $\{C_t\}$ in natural units. The value process of the hedging asset is $\{S_t\}$ in natural units. We assume that the hedging asset can be borrowed in any quantity at no cost, and that a short position in the hedging asset can be maintained and adjusted on a continuous basis at no cost. The value of the hedge portfolio at time $t$ is
    \begin{eqnarray} 
    V_t = C_t - \phi_t S_t + \theta_t \, ,
    \label{portfolio}
    \end{eqnarray}
    where the predictable process $\{\phi_t\}$ denotes the number of units of the hedging asset being shorted, and the predictable process $\{\theta_t\}$ denotes the number of benchmark units held in the hedge portfolio. Initially, we have $\theta_0 =\phi_0 S_0$. That is to say, the proceeds of the initial short sale of the hedging asset are deposited in the benchmark account.  Thereafter, the portfolio is managed on a self-financing basis: thus, the change in the value of the portfolio over a small interval of time is given by 
    \begin{eqnarray} 
    \rd V_t = \rd C_t - \phi_t \, \rd S_t \, .
    \label{error}
    \end{eqnarray}
    It follows from  \eqref{portfolio} and \eqref{error}  that the position in the benchmark account at time $t$ is 
    \begin{eqnarray} 
    \theta_t = \phi_t \,S_{t^-} - \int_0^{t^-} \phi_u \, \rd S_u \, ,
    \label{theta integral 1}
    \end{eqnarray}
    \vspace{-0.10cm}
    or equivalently
    \vspace{-0.5cm}
    \begin{eqnarray} 
    \theta_t = \phi_t \,S_{t} \,- \int_0^{t} \phi_u \, \rd S_u \, ,
    \label{theta integral 2}
    \end{eqnarray}
    where the integrals on the right-hand sides \eqref{theta integral 1} and \eqref{theta integral 2} are understood as being over the intervals $[0, t)$ and $[0, t]$, respectively. Then for the dynamics of the hedge portfolio we have
    \begin{align}
    {\rm d}V_t =  \big(\sigma^{c}_t  \, C_{t^-} - \phi_t  \, \sigma_t \, S_{t^-} \big)\,{\rm d}W_t 
     +  \int_{|x|>0} \big(\Sigma^{c}_t(x)  \, C_{t^-} - \phi_t \,\Sigma_t(x) \, S_{t^-} \big ) \tilde{{N}}({\rm d}x, {\rm d}t) \, .
    \label{Portfolio dynamics}
    \end{align}

    Now, if both of the assets are driven purely by the Brownian motion, and there are no jumps, then a perfect hedge can be carried out in such a way that the value of the hedge portfolio is constant. In that case a short calculation shows that 
    \begin{eqnarray} 
    \phi_t  = \frac {\sigma^{c}_t  \, C_{t} } { \sigma_t \, S_{t} }  \quad {\rm and} \quad
    \theta_t  = C_0 + \left( \frac {\sigma^{c}_t   } { \sigma_t  } - 1 \right) C_t \, .
    \end{eqnarray}
    The expression for the hedge ratio will look familiar, of course, but one should keep in mind that the hedge here is for the natural value of the contract asset, not its value in units of the fiat currency. In the general situation, when jumps are allowed, it is not possible to find a perfect hedge in the sense of completely erasing the riskiness of the position. Instead, we proceed as follows. We assume that the natural values of the assets under consideration are square-integrable in the sense that
    \begin{eqnarray} 
    \mathbb{E}\left[S_t^{\,2}\right] < \infty,   \quad \mathbb{E}\left[S_t \,C_t \right] < \infty, \quad  \mathbb{E}\left[C_t^{\,2}\right] < \infty, 
    \label{conditions on asset and contract asset}
    \end{eqnarray}
    for $t \geq 0$, and that the self-financing hedging strategy $\{ \phi_t, \theta_t \}_{t\geq 0}$ is such that the portfolio value at any time $t\geq0$ over which the hedge is maintained satisfies
    \begin{eqnarray} 
    \mathbb{E}\left[V_t^{\,2}\right] < \infty\,.
    \label{condition on value process}
    \end{eqnarray}
    Again, these assumptions are reasonable from a financial point of view, since we would not wish to consider assets that fail to satisfy such conditions as suitable for trading on a commercial basis. We fix a time interval $[0, \,T]$. Our goal is to choose the hedging strategy so as to minimize the expected squared deviation of the value of the hedge portfolio at time $T$ from its value at time $0$. Note that once we specify the positions held in the risky assets, the self-financing condition determines the corresponding holding required in the benchmark asset. Thus, if for any admissible choice of the strategy 
    $\Phi = \{\phi_t\}_{0 \leq t \leq T}$ we write
    \begin{eqnarray} 
    \Delta_T({\Phi}) = \mathbb{E}\left[(V_T - V_0)^2\right] 
    \end{eqnarray}
    for the corresponding mean squared error, then we have 
    \begin{align} 
     \Delta_T (\Phi)= \mathbb{E}\left[ \left( \int_{0}^{T} \big(\sigma^{c}_u C_{u^-} -\phi_u \sigma_u S_{u^-}\big) \,{\rm d}W_u  + 
    \int_{0}^{T}\int_{|x|>0} \big(\Sigma^{c}_u(x) C_{u^-} -\phi_u \Sigma_u(x) S_{u^-}\big) \, \tilde{{N}}({\rm d}x, {\rm d}u) \right
    )^2 \, \right]. \nonumber 
    \end{align}
    Therefore, by use of Proposition \ref{Proposition Ito isometry} we obtain
    \begin{eqnarray} 
    \Delta_T (\Phi)= \mathbb{E} \left[ \int_{0}^{T}\big(\sigma^{c}_u C_{u^-} -\phi_u \sigma_u S_{u^-}\big)^2 {\rm d}u  + \int_{0}^{T} \int_{|x|>0} \big(\Sigma^{c}_u(x) C_{u^-} -\phi_u \Sigma_u(x) S_{u^-}\big)^2 \, \nu({\rm d}x) \,{\rm d}u  \right]. \nonumber 
    \end{eqnarray}
    It follows that the mean squared error takes the form
    \begin{eqnarray} 
    \Delta_T (\Phi)= \mathbb{E} \left[ \int_{0}^{T} \left( K_u \,C_{u^-}^{\,2} - 2\phi_u \, L_u \, S_{u^-} C_{u^-}+  \phi_u^{\,2} M_u \, S_{u^-}^{\,2}\right) \,{\rm d}u   \right], 
    \end{eqnarray}
    where
    \begin{eqnarray} 
    K_t = \sigma^{c\,2}_t + \int_{x} {\Sigma^c_t(x)} ^2 \, \nu({\rm d}x), \,\,\,
    L_t = \sigma_t \,  \sigma^{c}_t  + \int_{x} \Sigma_t(x) \, \Sigma^{c}_t(x)  \, \nu({\rm d}x), \,\,\,
    M_t = \sigma_t ^{\,2 }+ \int_{x} \Sigma_t(x) ^2 \, \nu({\rm d}x)\,. \nonumber 
    \end{eqnarray}
    \noindent Thus we are led to the following.
    \vspace{0.2 cm}
    \begin{Proposition} Let $\{C_t\}$ be hedged with $\{\phi_t\}$ units of 
    $\{S_t\}$ and $\{\theta_t\}$ units of the benchmark.  Then the optimal hedge $\{\hat \phi_t, \hat \theta_t  \}_{0 \leq t \leq T}$ is given {\rm a.e.}-\,$\{\mathbb P \times {\rm Leb}\,[0,T]\}$ by
    \begin{eqnarray} 
    \hat{\phi}_t  = \frac{\sigma_t \,  \sigma^{c}_t  + \int_{x} \Sigma_t(x) \, \Sigma^{c}_t(x)  \, \nu({\rm d}x) }{\sigma_t ^{\,2} + \int_{x} \Sigma_t(x) ^2 \, \nu({\rm d}x)} \,  \frac{\,C_{t^-}} {\,S_{t^-}} \,, \quad
       \hat \theta_t = \hat \phi_t \,S_{t} \,- \int_0^{t} \hat \phi_u \, \rd S_u \, .
    \label{Optimal Hedge}
    \end{eqnarray}
    \end{Proposition}
    \vspace{0.1 cm}
    \proof
    A standard argument using the calculus of variations establishes \eqref{Optimal Hedge} as a candidate for the optimal hedge. The non-degeneracy condition imposed on the hedging asset ensures that the denominator is non-vanishing on $\{\Omega \times [0,T]\} \backslash \mathscr D$. To prove that the candidate is indeed optimal,  we need to show that the mean squared error in any alternative hedge is no less than the mean squared error in the candidate hedge. Let $\Psi = \{\psi_t\}_{0 \leq t \leq T}$ denote an alternative hedge. We say that two strategies $\{\psi_t^1\}$ and $\{\psi^2_t\}$ over $[0, T]$ are distinct if 
    \begin{eqnarray} 
    \mathbb P\times {\rm Leb}\,[0,T] \left [ \big( \psi_t^1 - \psi^2_t \big)^2 >0 \right ] >0 \, .
    \end{eqnarray}
A calculation then gives
    \begin{eqnarray} 
    \Delta_T \big(\Psi\big) - \Delta_T \big(\hat \Phi\big) 
    = \mathbb{E}\left[\int_{0}^{T} (\psi_u - \hat\phi_u)^2  \,
    S^{\,2}_{u^-}  \, M_u\,  
    {\rm d}u  \right ],
    \label{optimality inequality} 
    \end{eqnarray}
    where $\hat \Phi = \{\hat \phi_t\}_{0 \leq t \leq T}$, and one sees that the right hand side is nonnegative for any  alternative hedge. In fact, the optimal hedge dominates any distinct alternative. 
    \endproof
    We remark, incidentally, that one can substitute the optimal hedging strategy 
    \eqref{Optimal Hedge} back into the hedge portfolio value process $\{V_t\}$ with dynamics 
    \eqref{error} to check that the condition \eqref{condition on value process} is satisfied. In fact, one can use \eqref{optimality inequality} as a shortcut to get this result. If we compare the error when no hedge is put on to the error when the optimal hedge is put on, then we have
    \begin{eqnarray} 
   \Delta_T \big(\hat \Phi\big) = \Delta_T \big(0) -  \mathbb{E}\left[\int_{0}^{T}  \hat\phi_u^{\,2}  \,
    S^{\,2}_{u^-}  \, M_u\,  
    {\rm d}u  \right ],
    \label{no hedge} 
    \end{eqnarray}
    which gives a bound on  $\Delta_T \big(\hat \Phi\big)$. We just need to check that the terms on the right hand side of this relation are both finite. But
    \begin{eqnarray} 
    \Delta_T (0)= \mathbb{E} \left[ \int_{0}^{T} \bigg( {\sigma^{c}_u} ^{\,2}   +  \int_{|x|>0} \Sigma^{c}_u(x) ^{\,2} \, \nu({\rm d}x) \bigg)  C_{u^-}^{\,2} \,{\rm d}u  \right], 
    \end{eqnarray}
    which is finite on account our assumption that $C_T$ is square integrable. Then since  we assume that $S_T$ is square integrable and  that $S_TC_T$ is integrable, for the second term we get
    \begin{eqnarray} 
    \mathbb{E}\left[\int_{0}^{T}  \hat\phi_u^{\,2}  \, S^{\,2}_{u^-}  \, M_u  \, {\rm d}u  \right ] =
      \mathbb{E} \left[ \int_{0}^{T} \rho_u \, \bigg( {\sigma^{c}_u} ^{\,2}   +  \int_{|x|>0} \Sigma^{c}_u(x) ^{\,2} \, \nu({\rm d}x) \bigg)  \,
      C_{u^-}^{\,2} \,{\rm d}u  \right],
     \end{eqnarray}
    where
    \begin{eqnarray} 
    \rho_u = \frac{\bigg(\sigma_u \, \sigma^{c}_u    +  \int_{|x|>0} \Sigma_u(x) \, \Sigma^{c}_u(x)   \, \nu({\rm d}x) \bigg)^2}{\bigg( {\sigma_u} ^{\,2}   +  \int_{|x|>0} \Sigma_u(x) ^{\,2} \, \nu({\rm d}x) \bigg)
    \bigg( {\sigma^{c}_u} ^{\,2}   +  \int_{|x|>0} \Sigma^{c}_u(x) ^{\,2} \, \nu({\rm d}x) \bigg)}\,.
     \end{eqnarray}
   By virtue of the Cauchy-Schwartz inequality we have $0 \leq \rho_u \leq 1$, and this allows to conclude that the second term is also finite.  
    %
    \section{Optimal Hedging with Multiple Hedging Assets}
    %
    \noindent Let us now consider the more general problem of setting up a trading strategy to hedge the natural value of a position in a given contract asset  $\{C_t\}$ with a collection of $n$ hedging assets $\{S^i_t\}_{i = 1,...,n}$ each with dynamics 
    of the form \eqref{Natural Asset price bar drop}. Thus we have
    \begin{eqnarray}
    \frac{{\rm d} S^i_t}{\,\,\, S^i_{t^-}} = \sigma^{i}_t \,  {\rm d}W_t + \int_{|x|>0}  \Sigma^{i}
    _t(x) \, \tilde{{N}}({\rm d}x, {\rm d}t)\,. 
    \label{multiple asset dynamics}
    \end{eqnarray} 
    We shall assume the collection $\{S^i_t\}_{i = 1,...,n}$ is non-degenerate in the sense that no one of the assets can be replicated by holding a portfolio  in the remaining $n-1$ assets along with a position in the benchmark. More precisely, let the degeneracy subset $\mathscr D \subset \{\Omega \times [0, T]\}$  be the collection of points defined by 
      \begin{eqnarray}
 \mathscr D = \bigg\{ \omega \in \Omega, \, t \in [0, T] \, \bigg| \, \exists \, \iota^i_{t} \,: \,\sum_{i=1}^n    \iota^i_{t} \, \sigma^i_t  \,S^i_{t^-} =0 \, \cap \, \sum_{i=1}^n  \iota^i_{t} \, \Sigma^i_t (x) \, S^i_{t^-} =0, \, x \in S_\nu \bigg\}\,.
  \end{eqnarray} 
    If follows that on the complement $\mathscr D_c :=  \{\Omega \times [0, T]\} \backslash \mathscr D$, no self-financing trading strategy $\{\{ \iota^i_t \}_{i = 1, \dots, n}, \zeta_t\}$ in the $n$ hedging assets and the benchmark exists such that
    \begin{eqnarray}
    \sum_{i=1}^n    \iota^i_{t} \, \sigma^i_t  \,S^i_{t^-} =0\,  \quad {\rm and} \quad  \sum_{i=1}^n  \iota^i_{t} \, \Sigma^i_t (x) \, S^i_{t^-} =0 \, ,
    \end{eqnarray} 
  for $x\in \mathbb R^n$ in the support of the L\'evy measure. Our assumption is that the degeneracy subset should have $\{ \mathbb P \times {\rm Leb} \,[0,T] \}$-measure zero. We assume further that
    \begin{eqnarray} 
    \mathbb{E}\left[S^i_t \,S^j_t\right] < \infty,   \quad \mathbb{E}\left[S^i_t \,C_t \right] < \infty, \quad  \mathbb{E}\left[C_t^{\,2}\right] < \infty, 
    \label{square integrability conditions on assets and contract asset}
    \end{eqnarray}
for $t \in [0,T]$, $i,j = 1, \dots, n$. When $n$ such risky assets with negligible degeneracy are available for hedging, the hedge portfolio for the contract asset takes the form
    \begin{eqnarray}
    V_t = C_t - \sum_{i=1}^n \phi^i_t \, S^i_t \, + \theta_t \, ,
    \end{eqnarray} 
    and we impose the self-financing condition
    \begin{eqnarray}
    {\rm d}V_t = {\rm d}C_t - \sum_{i=1}^n \phi^i_t \, {\rm d}S^i_t \, .
    \label{multiple asset self financing condition}
    \end{eqnarray} 
    Our goal is to choose the hedging strategy $\Phi = \{\phi^i_t\}_{0 \leq t \leq T}$ in such a way that the mean squared error in the portfolio value
    \begin{eqnarray}
    \Delta_T(\Phi) = \mathbb{E}\, \left[(V_T - V_0)^2 \right]
    \label{Delta}
    \end{eqnarray} 
    is minimized. Then by \eqref{Second Natural Asset price bar drop}, \eqref{multiple asset dynamics}, \eqref{multiple asset self financing condition}  and \eqref{Delta} we have
    \begin{eqnarray} 
    \hspace{-14cm} \Delta_T(\Phi) =  \nonumber 
    \end{eqnarray}
    \vspace{-0.75cm}
    \begin{eqnarray} 
     \mathbb{E}\left[ \left( \int_{0}^{T} \big(\sigma^{c}_u C_{u^-} - \sum_{i=1}^n \phi^i_u \sigma^i_u S^i_{u^-}\big) \,{\rm d}W_u  + 
    \int_{0}^{T}\int_{|x|>0} \big(\Sigma^{c}_u(x) C_{u^-} - \sum_{i=1}^n \phi^i_u \Sigma^i_u(x) S^i_{u^-}\big) \, \tilde{{N}}({\rm d}x, {\rm d}u) \right
    )^2\right], \nonumber
    \end{eqnarray}
    and by use of the Ito isometry we obtain 
    \begin{eqnarray} 
    \hspace{-14cm} \Delta_T(\Phi) = \nonumber
    \end{eqnarray}
    \vspace{-0.75cm}
    \begin{eqnarray}
    \hspace{-0.4cm}\mathbb{E}\left[  \int_{0}^{T} \big(\sigma^{c}_u C_{u^-} - \sum_{i=1}^n \phi^i_u \sigma^i_u S^i_{u^-}\big)^2 \,{\rm d}u  + 
    \int_{0}^{T}\int_{x} \big(\Sigma^{c}_u(x) C_{u^-} - \sum_{i=1}^n \phi^i_u \Sigma^i_u(x) S^i_{u^-}\big)^2 \, \nu({\rm d}x) {\rm d}u \right]. \nonumber
    \end{eqnarray}
    Expanding the squares and gathering together the various terms we get
    \begin{eqnarray}
    \Delta_T(\Phi) = \mathbb{E}\left[\int_0^{T} \left(G_u + \sum_{i=1}^n \sum_{j=1}^n \phi^i_u \phi^j_u M^{ij}_u - 2 \sum_{i=1}^n \phi^i_u F^i_u\right)  \,{\rm d}u  \right] ,
    \label{Delta n assets}
    \end{eqnarray} 
    where
    \begin{eqnarray}
    M^{ij}_u = S^i_{u{^-}} S^j_{u{^-}}  \left[ \sigma^i_u \sigma^j_u  +  \int_{x} \Sigma^i_u(x) \Sigma^j_u(x) \nu({\rm d}x)  \right] ,
    \label{Mij}
    \end{eqnarray} 
    \begin{eqnarray}
    F^i_u = S^i_{u{^-}}  C_{u{^-}}  \left[ \sigma^i_u \sigma^c_u  +  \int_{x} \Sigma^i_u(x) \Sigma^c_u(x) \nu({\rm d}x) \right] ,
    \end{eqnarray} 
    \begin{eqnarray}
    G_u = C_{u{^-}}^{\,2} \left[ \sigma^{c\,2}_u + \int_{x}  {\Sigma^c_u}(x)^2 \nu({\rm d}x)\right].
    \end{eqnarray} 
    Applying a perturbation 
    \begin{eqnarray}
    \{\phi^i_t\}_{0 \leq t \leq T} \to \{ \phi^i_t + \epsilon^i_t \}_{0 \leq t \leq T}
    \end{eqnarray} 
    to the hedging strategy, we find that  the difference in the corresponding expressions for the mean squared errors is given by
    %
    %
    \begin{eqnarray}
    \Delta_T(\Phi + \epsilon) - \Delta_T(\Phi) = 
     \mathbb{E}\left[\int_0^{T}  \left(  \sum_{i=1}^n \sum_{j=1}^n \epsilon^i_u (\epsilon^j_u + 2 \phi^j_u)M^{ij}_u  - 2 \sum_{i=1}^n \epsilon^i_u F^i_u  \right)   {\rm d}u  \right] .
 \label{perturbation}
    \end{eqnarray} 
    A sufficient condition for the right-hand side of \eqref{perturbation} to vanish to first order in the perturbing variables, and hence lead to a candidate optimum,  is that the $\{\phi^i_t\}$ should satisfy 
    \begin{eqnarray}
    \sum_{j=1}^n M^{ij}_t \phi^j_t =   F^i_t \,, \quad {\rm a.e.}\text{-}\{\mathbb P \times {\rm Leb}\,[0,T]\} ,
        \label{1st order condition for n assets}
    \end{eqnarray} 
 and we are thus led to the following.
    \vspace{0.2cm}
    \begin{Proposition} Let $\{C_t\}$  be hedged with $\{\phi^i_t\}$ units of 
    $\{S^i_t\}$ for $i = 1, \dots, n$ and $\{\theta_t\}$ units of the benchmark.  Then  the optimal hedge  takes the form
    \begin{eqnarray}
    \hat{\phi}^i_t = \sum_{j=1}^n N^{ij}_t F^j_t \,, \quad
    \hat \theta_t = \sum_{i=1}^n \hat \phi^i_t \, S^i_t - \int_0^t \sum_{i=1}^n \hat \phi^i_u \, {\rm d}S^i_u \, , 
    \quad {\rm a.e.}\text{-}\,\{\mathbb P \times {\rm Leb}[0,T]\} \, ,
    \label{optimal hedge n assets}
    \end{eqnarray} 
   where $\{N^{ij}_t\}$ is the inverse of $\{M^{ij}_t\}$ on $\mathscr D_c$.
    \end{Proposition}
    \vspace{0.2cm}
    \proof
    The inverse of the matrix $\{M^{ij}_t\}$ exists on  $\mathscr D_c$ on account of the non-degeneracy condition that we have imposed on the collection of hedging assets. In particular, it follows from the definition of $\mathscr D$ that $\{\omega, t\} \in \mathscr D_c$
    if and only if the inequality
    \begin{eqnarray}
     \left(\sum_{i=1}^n  \iota^i_t \sigma^i_u S^i_{u{^-}}\right)^2  +  \int_{x} \left(\sum_{i=1}^n  \iota^i_t \, \Sigma^i_u(x)  S^i_{u{^-}}\right)^2   \nu({\rm d}x)  > 0
    \end{eqnarray} 
holds for any non-vanishing hedging strategy $\{\iota^i_t\}_{i=1,...,n}$. But this relation is equivalent to 
    \begin{eqnarray}
    \sum_{i=1}^n \sum_{j=1}^n \, M^{ij}_t \, \iota^{j}_{t} \, \iota^{i}_{t} > 0\,,
    \label{positive definite}
    \end{eqnarray} 
which shows that  $\{M^{ij}_t \}$  is positive definite on $\mathscr D_c$, and hence possesses an inverse. The solution of equation \eqref{1st order condition for n assets} then gives a candidate optimal hedge. As in the case of a single hedging asset, we need to show that the error in any alternative hedge is no less than the error in the candidate solution. Putting \eqref{optimal hedge n assets} back into \eqref{Delta n assets}, we get
    \begin{eqnarray}
    \Delta_T \big(\hat \Phi \big) = \mathbb{E}\left[\int_0^{T} \left(G_u - \sum_{i=1}^n \sum_{j=1}^n  N_u^{ij} F^{j}_u F^{i}_u\right) {\rm d}u  \right] .
    \end{eqnarray} 
    Then, letting $\{\psi^i_t\}_{0 \leq t \leq T}$ be any alternative hedge that is distinct from the candidate \eqref{optimal hedge n assets},  one finds that
    \begin{eqnarray}
    \Delta_T\big(\Psi \big) - \Delta_T\big(\hat \Phi \big)  =  \mathbb{E}\left[\int_0^{T} \sum_{i=1}^n \sum_{j=1}^n  M_u^{ij} \, (\psi^i_u - \hat{\phi}^i_u)(\psi^j_u - \hat{\phi}^j_u)\,{\rm d}u  \right] .
    \label{Delta psi - Delta phi n assets}
    \end{eqnarray} 
    The right side of \eqref{Delta psi - Delta phi n assets} is strictly positive, and we deduce that $\{\hat{\phi}^i_t\}$ is optimal and indeed that it dominates any strategy distinct from it.
    \endproof
    \vspace{0.2cm}
    Next, we wish to show that if we add a further non-redundant hedging asset to an existing collection of $n$ hedging assets satisfying a non-degeneracy condition, the hedge will be improved by using all $n+1$ of the hedging assets. This is a characteristic feature of incomplete markets. Given $\{C_t\}$ and $\{S^i_t\}_{i = 1,...,n}$, let $\{\hat{\phi}^i_t\}_{i = 1,\dots,n}$ denote the optimal hedge determined in Proposition 5, and let $\{S^0_t\}$ be another hedging asset, which is taken to be non-redundant in the sense that it cannot be realized as a portfolio formed from the original $n$ hedging assets together with the benchmark asset. 
    
    More precisely, let us now write $\mathscr D^n$ (in place of $\mathscr D$) for the degeneracy set associated with the $n$ original hedging instruments, which we have assumed to be of $\{\mathbb P \times {\rm Leb}(0,T)\}$-measure zero, and let us write $\mathscr D^{n+1}$ for the degeneracy set of the enhanced collection of $n+1$ heading instruments, which we also assume to have $\{\mathbb P \times {\rm Leb}(0,T)\}$-measure zero. It should be evident that $\mathscr D^n \subset \mathscr D^{n+1}$, since there may be points in $\{\Omega \times [0, T] \}$ at which the enhanced collection degenerates even though the original collection is non-degenerate. Then we have the following.
    
    \vspace{0.2cm}
    \begin{Proposition}For any contract asset $\{C_t\}$, the optimal hedge $\{\hat{\Gamma}^i_t\}_{i = 0,1,\dots,n}$ obtained by use of the enhanced collection of $n+1$ hedging assets $\{S^i_t\}_{i = 0,1,\dots,n}$ is strictly better than the optimal hedge $\{\hat{\phi}^i_t\}_{i =1,\dots,n}$ obtained by use of the original $n$ hedging assets $\{S^i_t\}_{i = 1,\dots,n}$.  
    \end{Proposition}
    \proof
    The argument proceeds in two steps. First, let  $\{\hat U_t\}_{0 \leq t \leq T}$ denote the value process of the optimal hedge position 
    $\{\hat \phi^i_t, \hat \theta_t \}$ constructed from the original $n$ hedging assets together with the benchmark asset, as determined in Proposition 5. Thus, we have
    \begin{eqnarray}
    \hat U_t = \sum_{i=1}^n \hat \phi^i_t \, S^i_t + \hat \theta_t  \,,
    \end{eqnarray} 
    where $\{ \hat \phi^i_t, \hat \theta_t\}$ is given as in  \eqref{optimal hedge n assets}. It follows by the self-financing condition that $\{\hat U_t\}$ itself can be treated as an asset. Now consider a hedging strategy of the form $\{\gamma_t, \delta_t, \epsilon_t \}$ where $\{\gamma_t\}$ denotes the holdings in $\{U_t\}$, where $\{\delta_t\}$ denotes the holdings in $\{S^0_t\}$, and $\{\epsilon_t\}$ denotes the holdings in the benchmark asset. It is easy to see that an optimal hedge involving a pair of non-redundant risky hedging instruments will perform better than the optimal hedge obtained by use of just one of the two risky instruments. This is because the optimal hedge involving a single risky instrument is an example of a sub-optimal hedge involving two risky instruments. It follows that as a hedge for $\{C_t\}$ the strategy $\{\gamma_t, \delta_t, \epsilon_t \}$ with value process $\{\gamma_t U_t + \delta_t S^0_t + \epsilon_t \}$ will perform strictly better than the strategy $ \{ 1, \hat \theta_t \}$ with value process $\{U_t + \hat \theta_t \}$ . That is to say, 
    \begin{eqnarray}
    \Delta_T \big(\gamma \, \hat{\Phi}, \delta \big) < \Delta_T\big(\hat{\Phi}, 0\big) \, .
    \label{two risky assets}
    \end{eqnarray} 
    On the other hand, we observe that if $\hat \Gamma := \{ \hat{\Gamma}^i_t\}_{i = 0,1,...,n}$ denotes the {\em optimal} enhanced hedging strategy  involving the $n+1$ assets now available for hedging, along with a position $\{\zeta_t\}$ in the benchmark, then the portfolio $\{\gamma_t \, \hat{\phi}_t^i, \delta_t, \epsilon_t\}_{i = 1,...,n}$ considered above  at \eqref{two risky assets} is merely an example of a hedge involving the $n+1$ hedging assets, and though it might be optimal, in general it will be suboptimal.  Therefore
    \begin{eqnarray}
    \Delta_T\big(\hat \Gamma  \big) \leq \Delta_T\big(\gamma \hat{\Phi}, \delta \big),
    \end{eqnarray} 
    \vspace{-0.2cm}
    \hspace{-0.25cm} and hence
    \vspace{-0.2cm}
    \begin{eqnarray}
    \Delta_T\big(\hat \Gamma \big)< \Delta_T\big(\hat{\Phi}, 0\big)  .
    \end{eqnarray} 
    It follows that the optimal hedge involving $n+1$ hedging instruments will perform better than the optimal hedge formed from any $n$ of them. 
    \endproof
    
    \section{Simulations}
    %
    \noindent
    In conclusion, we propose in this section to look in more detail at the $n=2$ case and consider simulating the optimal trading strategy to hedge the natural value of a position in a given contract asset by use of two risky hedging assets. The problem will be framed in the case where all three of the assets are driven by a one-dimensional Brownian motion $\{W_t\}$ and an independent one-dimensional Poisson random measure $\{{N}({\rm d}x, {\rm d}t)\}$. The hedging assets each have dynamics of the form \eqref{Natural Asset price bar drop}. We write $\{S^{i}_t\}_{i=1, \,2}$ for the hedging assets, and we write $\{\phi^{i}_t\}_{i=1, \,2}$ for the holdings in these assets. Then the rather general construction given in Section V leads to the following:
    \vspace{0.2 cm}
    \begin{Proposition} Let the contract asset $\{C_t\}$ be hedged over $[0, T]$ with $\{\phi^{1}_t\}$ units of 
    $\{S^1_t\}$, $\{\phi^{2}_t\}$ units of 
    $\{S^2_t\}$, and $\{\theta_t\}$ units of the benchmark.~Then the optimal hedge  is given by 
    	\begin{eqnarray} 
    	\hat\phi^{1}_t  =   \frac{P^{12}_t-Q^{12}_t}{R^{12}_t} \, \frac{\,C_{t^{-}}}{\,S^1_{t^{-}}}, \quad \quad
    	\hat\phi^{2}_t  =   \frac{P^{21}_t-Q^{21}_t}{R^{21}_t} \, \frac{\,C_{t^{-}}}{\,S^2_{t^{-}}},
    	\label{Optimal Hedge for two assets}
    	\end{eqnarray}
    	on the non degeneracy subset $\mathscr D_c$, where we write
    	%
    	\begin{eqnarray} 
    	P^{ij}_t = \left(\sigma^c_t \, \sigma^i_t + \int_{x} \Sigma^c_t(x) \, \Sigma^i_t(x) \, \nu({\rm d} x)\right)\left( {\sigma^{j\,2}_t}+ \int_{x} {\Sigma^j_t}(x)^2 \, \nu({\rm d} x) \right), \nonumber
    	\end{eqnarray}
    	\vspace{-0.3 cm}
    	\begin{eqnarray} 
    	Q^{ij}_t = \left(\sigma^i_t \, \sigma^j_t + \int_{x} \Sigma^i_t(x) \,\Sigma^j_t(x) \,\nu({\rm d} x)\right)\left(\sigma^c_t \,\sigma^j_t + \int_{x} \Sigma^c_t(x)\, \Sigma^j_t(x) \,\nu({\rm d} x)\right), \nonumber
    	\end{eqnarray}
    	\vspace{-0.3 cm}
    	\begin{eqnarray} 
    	R^{ij}_t =  \left(\sigma^{i \,2}_t + \int_{x} {\Sigma^i_t}(x)^2 \,\nu({\rm d} x)\right) 
    	\left(\sigma^{j\,2}_t + \int_{x} {\Sigma^j_t} (x)^2 \nu({\rm d} x)\right) - \left(\sigma^i_t \, \sigma^j_t + \int_{x} \Sigma^i_t(x) \, \Sigma^j_t(x) \nu({\rm d} x)\right)^2. \nonumber
    	\end{eqnarray}
    \end{Proposition}
    
    \vspace{0.4cm}

    As a numerical illustration of the general methodology let us consider the situation where each of the assets follows a geometric 
    L\'evy process for which the L\'evy process takes the form of a jump diffusion consisting of a standard Brownian motion superposed on  a compound Poisson process. It should be recalled that even if the driving process  in the exponent  of the asset price is a L\'evy process, the asset price itself follows a L\'evy-Ito process. 
    
    We consider the simplest possible case, namely, that for which the pure-jump component of the L\'evy process is a Bernoulli process. Let $\{X_t\}_{t \geq 0}$ denote a compound Poisson process for which the jumps arrive randomly according to a Poisson process $\{N_t\}_{t \geq 0}$ with rate $m$. The jump sizes  
    $\{Y_i\}_{i \in \mathbb N}$ are independent identically-distributed random variables. We assume that  $\{Y_i\}_{i \in \mathbb N}$ and $\{N_t\}$ are independent. Let us write $Y$ for a typical element of the set $\{Y_i\}_{i \in \mathbb N}$. In the example under consideration we shall assume that $Y$ has a Bernoulli distribution ${\rm Bern}(g,h; p)$. Thus $Y$ takes values in a set  $\{g, h\}$ where 
    $g,h \in \mathbb R$ with $\mathbb P [Y = g] = p$ and $\mathbb P [Y = h] = 1-p$ . The L\'evy measure for such a process $\{X_t\}$ takes the form
    \begin{eqnarray}
    \nu({\rm d}x) = m\,\big(p \delta_g({\rm d}x) + (1-p) \delta_h({\rm d}x)\big)\, ,
    \end{eqnarray} 
    where  $\delta_g({\rm d}x)$ is the Dirac measure concentrated at $g$ and $\delta_h({\rm d}x)$ is the Dirac measure concentrated at $h$. Then the price processes of the assets under consideration have dynamics of the form 
    \eqref{Natural Asset price bar drop}-\eqref{Second Natural Asset price bar drop}, with deterministic time-independent volatilities. Since we are working with a geometric L\'evy process, the jump volatility is of the form $\Sigma(x) =  \exp (\beta x) - 1$, for some $\beta \in \mathbb{R^{+}}$. The price of a typical non-dividend paying risky asset in a Bernoulli jump diffusion market with this set up is thus of the form
    \begin{eqnarray}
    S_t = S_0 \, \exp\left(\sigma W_t - \frac{1}{2}\sigma^2 t + \beta X_t - mt\, \big( \, p\,(\re^{\beta\, g } - 1) + (1-p)\, (\re^{\beta \,h} - 1) \,\big) \right) ,
    \label{price process for Bernoulli}
    \end{eqnarray} 
    \noindent where $\sigma$ is a constant. For our simulations we consider a contract asset $\{C_t\}$ and a pair of hedging  assets $\{S^1_t\}$ and $\{S^2_t\}$, each of the form \eqref{price process for Bernoulli}, with a view to  forming an optimal hedge of the contract asset with positions in one or both of the hedging assets. 
    
\begin{figure}[ht!]
        \begin{minipage}[t]{0.5\textwidth}
            \centering
            \includegraphics[height=2.1in]{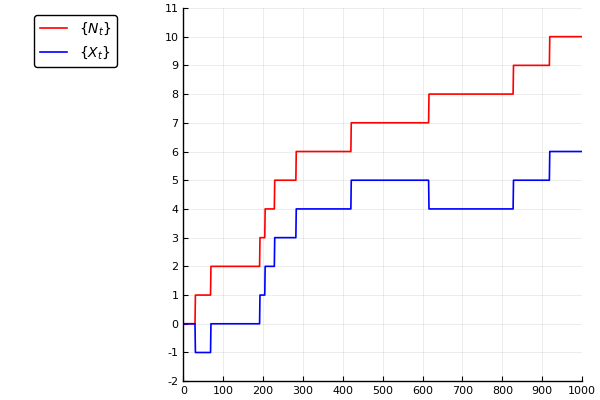}
        \end{minipage}%
        \begin{minipage}[t]{0.5\textwidth}
            \centering
            \includegraphics[height=2.1in]{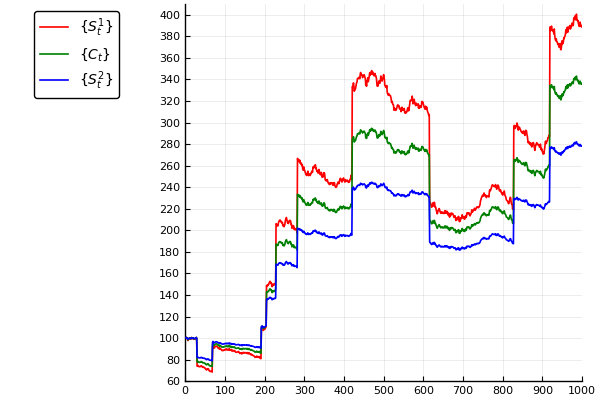}
        \end{minipage}
        \vspace{-5pt}
        \caption{ {\em Bernoulli jump-diffusion market.} The chart on the left above shows an outcome of chance for the L\'evy process in blue, with the underlying Poisson process in red.  The chart on the right above plots the value process of the contract asset in green.  The high volatility hedging asset 1 is shown in red, and the low volatility hedging asset 2 is shown in blue.}
    \label{fig: 1}
    \end{figure}

    In Figure \ref{fig: 1}, we show on the left-hand side a random sample path for the L\'evy process $\{X_t\}$ alongside the underlying Poisson process $\{N_t\}$. On the right-hand side one finds the corresponding paths for the contract asset $\{C_t\}$ and the two hedging assets $\{S^1_t\}$ and $\{S^2_t\}$. The inputs for this example are as follows: $S^1_0 = 100$, $S^2_0 = 100$, $C_0 = 100$, $\sigma^1 = 0.20$,  $\sigma^2 = 0.10$,  $\sigma^c = 0.15$, $\beta^1 = 0.30$, $\beta^2 = 0.20$, $\beta^c = 0.25$, $m = 15$, $p = 0.5$, $g = 1$, $h = -1$, and $T = 1$. The unit of time depicted on the $x$-axis is divided into a thousand parts.
    
\begin{figure}[ht!]
        \centering
        \begin{minipage}[t]{0.5\textwidth}
            \centering
            \includegraphics[height=2.1in]{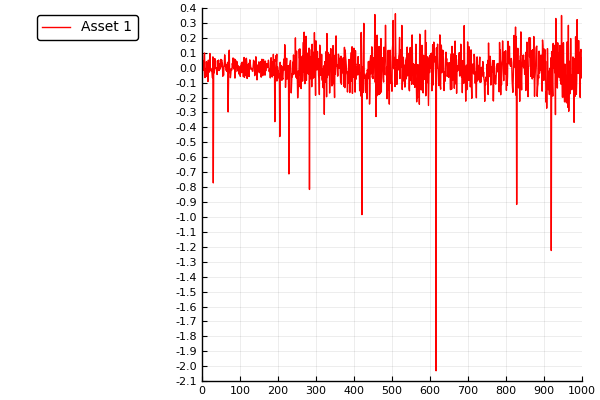}
        \end{minipage}%
        \begin{minipage}[t]{0.5\textwidth}
            \centering
            \includegraphics[height=2.1in]{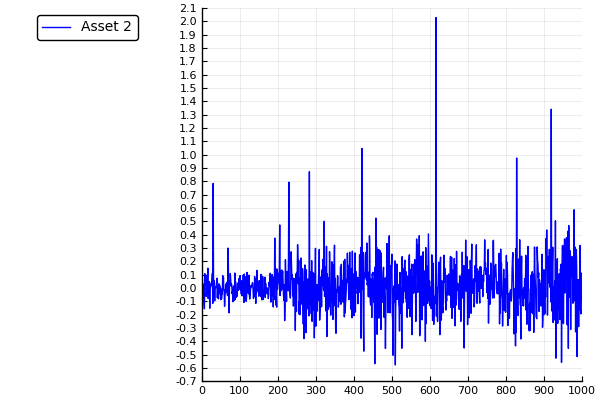}
        \end{minipage}
        \vspace{-5pt}
        \caption{ {\em Single-asset hedges.}  The chart on the left plots at each step the change in the value of the hedge portfolio, when asset 1 alone is used as the hedge. The lengthy downward spikes correspond to jumps, whereas the shorter spikes are due to Brownian volatility. In the chart on the right, asset 2 alone is used as the hedge. The lengthy upward spikes correspond to jumps.}
    \label{fig: 2}
    \end{figure}

    Now, we know from general theory that if the Brownian motion is non-vanishing then the hedge can never be perfect; but  if the Brownian component is small for all three assets, then a reasonably good hedge should be obtainable using just two assets in the case of a Bernoulli jump diffusion. In Figure \ref{fig: 2} we show the effect of using either $\{S^1_t\}$ or $\{S^2_t\}$ alone as a hedge and we plot the residual movements in the values of the hedged portfolios.

    \vspace{15pt}
    
    \vskip 2pt
    \newpage
    
\begin{figure}[ht!]
        \centering
            \includegraphics[height=2.2in]{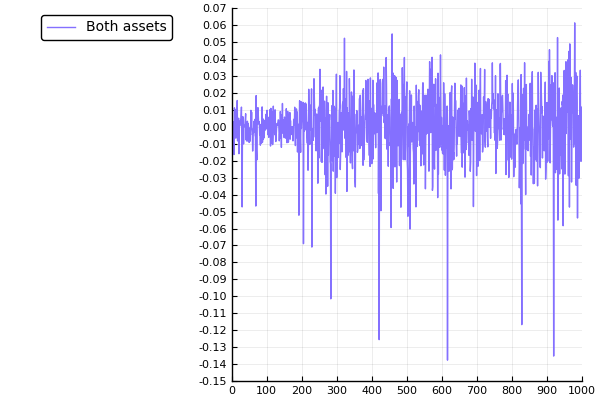}
        \vspace{3pt}
        \caption{ {\em Two-asset hedge.} The figure above plots the change in the value of the hedge portfolio when both hedging asset 1 and hedging asset 2 are included in the hedging strategy for the contract asset. The Brownian volatilities in this example are  $\sigma^1 = 0.20$,  $\sigma^2 = 0.10$ and $\sigma^c = 0.15$.}
    \label{fig: 3}
    \end{figure}

\begin{figure}[ht!]
        \centering
            \includegraphics[height=2.2in]{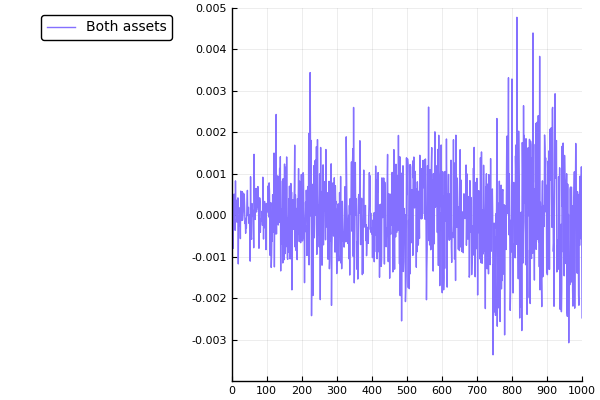}
        \vspace{3pt}
        \caption{ {\em Two-asset hedge with reduced Brownian volatilities.} This figure plots the change in the value of the hedge portfolio when both asset 1 and asset 2 are included in the hedging strategy for the contract, with $\sigma^1 = 0.003$,  $\sigma^2 = 0.001$ and $\sigma^c = 0.002$. In this example, a near-perfect hedge is obtained. Note that the scale of the $y$-axis is smaller than that of the previous figure.}
    \label{fig: 4}
    \end{figure}

     In Figure \ref{fig: 3} we show the effect of using both hedging assets together to hedge the contract asset, and we note in particular the significant drop in the variance of the hedged portfolio. If we reduce the volatilities of the Brownian components still further, then we get a near perfect hedge, as illustrated in Figure \ref{fig: 4}. The Brownian volatilities for Figure \ref{fig: 4} are given by $\sigma^1 = 0.003$,  $\sigma^2 = 0.001$ and $\sigma^c = 0.002$. 
    \vspace{1cm}
    
    \newpage
    It is sometimes said that L\'evy markets are incomplete except in the Brownian case, in the situation where the number of available assets is no less than the number of Brownian motions. But this of course is not quite true, since a pure Poisson market is also complete. If a pair of geometric L\'evy assets are driven by a common Poisson process, then either can be hedged by use of the other. A pure Bernoulli market is also complete, in the sense that if three geometric L\'evy assets are driven by a common Bernoulli process, then any one can be hedged by use of the other two. Similarly, a compound Poisson process market with $k$ possible outcomes at each jump is complete if $k$ hedging assets are available. If a Brownian component is introduced into any of these scenarios, then the resulting market is incomplete; but if the Brownian volatilities are small, then near perfect hedges can be achieved, as we see in Figure 4.

    \vskip 25pt 
    \noindent {\bf APPENDIX}\\
    
    \noindent
    Here we present some useful versions of the Ito product and quotient rules for L\'evy-Ito processes. The Brownian versions of these rules will be familiar, but the corresponding L\'evy-Ito rules do not seem previously to have been presented systematically in all their different versions, so we do so here.
     Let $\{X^1_t\}_{t \geq 0}$ and $\{X^2_t\}_{t \geq 0}$ be L\'evy-Ito processes, each satisfying dynamical equations of the form 
    \eqref{Levy Ito process differential form}, such that
    \begin{eqnarray} 
    {\rm d}X^1_t =  \alpha^1_t \, {\rm d}t +  \beta^1_t \, {\rm d}W_t + \int_{|x|\in(0,1)} \gamma^1_t(x) \, \tilde{{N}}({\rm d}x, {\rm d}t) + \int_{|x|\geq1} \delta^1_t(x) \, {N}({\rm d}x, {\rm d}t)\,
    \label{Levy Ito process differential form 1}
    \end{eqnarray}
    and
    \begin{eqnarray} 
    {\rm d}X^2_t =  \alpha^2_t \, {\rm d}t +  \beta^2_t \, {\rm d}W_t + \int_{|x|\in(0,1)} \gamma^2_t(x) \, \tilde{{N}}({\rm d}x, {\rm d}t) + \int_{|x|\geq1} \delta^2_t(x) \, {N}({\rm d}x, {\rm d}t)\,.
    \label{Levy Ito process differential form 2}
    \end{eqnarray}
    \vspace{0.2cm}
    \begin{Lemma} \label{product rule} The product rule for L\'evy-Ito processes takes the following form:
    \begin{align} 
    {\rm d}(X^1_t\,X^2_t) &= [\alpha^1_t X^2_{t^{-}} + \alpha^2_t X^1_{t^{-}} + \beta^1_t \beta^2_t]\,{\rm d}t  + (\beta^1_t X^2_{t^{-}}+ \beta^2_t X^1_{t^{-}}) {\rm d}W_t + \int_{|x|<1} \gamma^1_t(x)\, \gamma^2_t(x) \,\nu({\rm d}x)\,{\rm d}t \nonumber
    \\&
    + \int_{|x|\in(0,1)} (\gamma^1_t(x)\, \gamma^2_t(x) + \gamma^1_t(x)\, X^2_{t^{-}} + \gamma^2_t(x)\,X^1_{t^{-}}) \,\tilde{N}({\rm d}x, {\rm d}t)\nonumber
    \\&
    + \int_{|x| \geq 1} (\delta^1_t(x)\, \delta^2_t(x) + \delta^1_t(x)\, X^2_{t^{-}} + \delta^2_t(x)\,X^1_{t^{-}}) \,{N}({\rm d}x, {\rm d}t)\,.
    \label{Ito product rule}
    \end{align}
    \end{Lemma}
    \proof This is similar to the proof of the corresponding result for Ito processes, and is obtained by applying Ito's formula to each side of the identity 
    \begin{equation}
    X^1_t\,X^2_t = \frac{1}{4}\left( X^1_t + X^2_t \right)^2 - \frac{1}{4}\left( X^1_t - X^2_t \right)^2 .
    \end{equation}
    \noindent A calculation then gives the result claimed. 
    \endproof
    \vspace{0.2cm}
    Now let $\{X^1_t\}$ and $\{X^2_t\}$ be L\'evy-Ito processes such that $\{X^2_t\}$, $\{X^2_{t^-}\}$ are strictly positive. Then we obtain the following.
    \vspace{.2cm}
    \begin{Lemma} \label{quotient rule} The quotient rule for L\'evy-Ito processes is given by
    \begin{align} 
    {\rm d}\left(\frac{X^1_t}{X^2_t}\right) &= \left[\frac{\alpha^1_t X^2_{t^{-}} - \alpha^2_t X^1_{t^{-}}} {(X^2_{t^{-}})^2} + \frac{(\beta^2_t)^2 X^1_{t^{-}} - \beta^1_t \beta^2_tX^2_{t^{-}}}{(X^2_{t^{-}})^3} \right]{\rm d}t \nonumber
    \\&
    + \frac{\beta^1_t X^2_{t^{-}} - \beta^2_t X^1_{t^{-}}} {(X^2_{t^{-}})^2} {\rm d}W_t + \int_{|x| < 1} \frac{(\gamma^2_t(x))^2 X^1_{t^{-}} -\gamma^1_t(x)\gamma^2_t(x)X^2_{t^{-}}}{(X^2_{t^{-}})^2 (X^2_{t^{-}} + \gamma^2_t(x))} \nu({\rm d}x){\rm d}t \nonumber
    \\&
    +\int_{|x|\in(0,1)} \frac{\gamma^1_t(x)X^2_{t^{-}} - \gamma^2_t(x)X^1_{t^{-}}}{X^2_{t^{-}}(X^2_{t^{-}} + \gamma^2_t(x))} \tilde{N}({\rm d}x, {\rm d}t)  
    +\int_{|x| \geq 1} \frac{\delta^1_t(x)X^2_{t^{-}} - \delta^2_t(x)X^1_{t^{-}}}{X^2_{t^{-}}(X^2_{t^{-}} + \delta^2_t(x))} {N}({\rm d}x, {\rm d}t)\,. 
    \label{Ito quotient rule}
    \end{align}
    \end{Lemma}
    \proof First one uses Proposition 1 to work out the dynamics of the process $\{1/X^2_t\}$. Then one uses Lemma 1 to work out the dynamics of the product $\{ X^1_t \times 1/X^2_t\}$.  
    \endproof
    \vspace{0.2cm}
    For applications in finance, one often makes use of the ``proportional" versions of the L\'evy-Ito product and quotient rules, which are applicable if we assume that $\{X^1_t\}$, $\{X^1_{t^-}\}$, $\{X^2_t\}$, $\{X^2_{t^-}\}$ are strictly positive. The dynamical equations for $\{X^1_t\}$ and $\{X^2_t\}$ will be assumed in Lemmas 3 and 4 to take the proportional form 
    \begin{eqnarray} 
    {\rm d}X^1_t =  X^1_{t^-}\left[  \alpha^1_t \, {\rm d}t +  \beta^1_t \, {\rm d}W_t + \int_{|x|\in(0,1)} \gamma^1_t(x) \, \tilde{{N}}({\rm d}x, {\rm d}t) + \int_{|x|\geq1} \delta^1_t(x) \, {N}({\rm d}x, {\rm d}t) \right]\,
    \label{Levy Ito process differential form 1 proportional}
    \end{eqnarray}
    and
    \begin{eqnarray} 
    {\rm d}X^2_t =  X^2_{t^-}\left[  \alpha^2_t \, {\rm d}t +  \beta^2_t \, {\rm d}W_t + \int_{|x|\in(0,1)} \gamma^2_t(x) \, \tilde{{N}}({\rm d}x, {\rm d}t) + \int_{|x|\geq1} \delta^2_t(x) \, {N}({\rm d}x, {\rm d}t)\right] \,.
    \label{Levy Ito process differential form 2 proportional}
    \end{eqnarray}
    Then we have the following formulae, which arise as consequences of Lemmas 1 and 2. 
    \vspace{0.2cm}
    \begin{Lemma} \label{proportional product rule} In the proportional case, the product rule takes the form 
    \begin{align} 
    {\rm d}(X^1_t\,X^2_t) = X^1_{t^{-}} X^2_{t^{-}} & \bigg[ \,\left(\alpha^1_t + \alpha^2_t + \beta^1_t \,\beta^2_t + \int_{|x| < 1} \gamma^1_t(x)\, \gamma^2_t(x) \,\nu({\rm d}x)\,\right)\,{\rm d}t  + (\beta^1_t + \beta^2_t) {\rm d}W_t  \nonumber
    \\&
    + \int_{|x|\in(0,1)} \left(\gamma^1_t(x) \, \gamma^2_t(x) + \gamma^1_t(x) + \gamma^2_t(x) \right) \,\tilde{N}({\rm d}x, {\rm d}t)\nonumber
    \\& \quad
    + \int_{|x| \geq 1} \left(\delta^1_t(x) \, \delta^2_t(x) + \delta^1_t(x)+ \delta^2_t(x)  \right) \,{N}({\rm d}x, {\rm d}t) \,\bigg ]\,.
    \label{Ito proportional product rule}
    \end{align}
    %
    \end{Lemma}
    \vspace{.2cm}
    \begin{Lemma} \label{proportional quotient rule} In the proportional case, the quotient rule takes the form
    \begin{align} 
    &{\rm d}\left(\frac{X^1_t}{X^2_t}\right) = \frac{X^1_{t^{-}}}{X^2_{t^{-}}}\bigg[\left(\alpha^1_t-\alpha^2_t - \beta^2_t(\beta^1_t - \beta^2_t) - \int_{|x| < 1} \gamma^2_t(x) \frac{\gamma^1_t(x)-\gamma^2_t(x)}{1+\gamma^2_t(x)}\nu({\rm d}x)\right){\rm d}t  \nonumber 
    \\&
    + (\beta^1_t-\beta^2_t)\,{\rm d}W_t  + \int_{|x|\in(0,1)} \frac{\gamma^1_t(x) - \gamma^2_t(x)}{1+\gamma^2_t(x)}\tilde{N}({\rm d}x, {\rm d}t) + \int_{|x| \geq 1} \frac{\delta^1_t(x) - \delta^2_t(x)}{1+\delta^2_t(x)} {N}({\rm d}x, {\rm d}t)\bigg].
    \label{Proportional Quotient rule}
    \end{align}
    \end{Lemma}
    \vspace{0.2cm}
    The various forms of the Ito product and quotient rules simplify for symmetric proportional processes, and the resulting formulae are extremely useful in applications. Let $\{X^1_t\}$, $\{X^1_{t^-}\}$, $\{X^2_t\}$, $\{X^2_{t^-}\}$ be strictly positive  For symmetrical processes we can write 
    \begin{eqnarray} 
    {\rm d}X^1_t =  X^1_{t^-}\left[  \alpha^1_t \, {\rm d}t +  \beta^1_t \, {\rm d}W_t + \int_{|x|>0} \gamma^1_t(x) \, \tilde{{N}}({\rm d}x, {\rm d}t) \right]
    \label{Levy Ito process 1 proportional restricted}
    \end{eqnarray}
    and
    \begin{eqnarray} 
    {\rm d}X^2_t =  X^2_{t^-}\left[  \alpha^2_t \, {\rm d}t +  \beta^2_t \, {\rm d}W_t + \int_{|x|>0} \gamma^2_t(x) \, \tilde{{N}}({\rm d}x, {\rm d}t) \right] ,
    \label{Levy Ito process 2 proportional restricted}
    \end{eqnarray}
    and we obtain the following.
    \vspace{0.2cm}
    \begin{Lemma} \label{proportional symmetric product rule} In the symmetric proportional case the product rule takes the form 
    \begin{align} 
    {\rm d}(X^1_t\,X^2_t) = X^1_{t^{-}} X^2_{t^{-}}& \bigg[ \,\left(\alpha^1_t + \alpha^2_t + \beta^1_t \,\beta^2_t + \int_{x} \gamma^1_t(x)\, \gamma^2_t(x) \,\nu({\rm d}x)\,\right){\rm d}t  + (\beta^1_t + \beta^2_t) {\rm d}W_t \nonumber
    \\&\quad
    + \int_{|x|>0} \left(  \gamma^1_t(x) + \gamma^2_t(x) + \gamma^1_t(x) \, \gamma^2_t(x) \right) \,\tilde{N}({\rm d}x, {\rm d}t)
    \bigg ] .
    \label{symmetric proportional product rule}
    \end{align}
    \end{Lemma}
    \vspace{0.2cm}
    \begin{Lemma} \label{proportional symmetric quotient rule} In the symmetric proportional case the quotient rule takes the form 
    \begin{align} 
    {\rm d}\left(\frac{X^1_t}{X^2_t}\right) = \frac{X^1_{t^{-}}}{X^2_{t^{-}}} & \bigg[\left(\alpha^1_t-\alpha^2_t - \beta^2_t(\beta^1_t - \beta^2_t) - \int_{x} \gamma^2_t(x) \frac{\gamma^1_t(x)-\gamma^2_t(x)}{1+\gamma^2_t(x)}\nu({\rm d}x)\right){\rm d}t  \nonumber 
    \\&\quad
    + (\beta^1_t-\beta^2_t)\,{\rm d}W_t  + \int_{|x|>0} \frac{\gamma^1_t(x) - \gamma^2_t(x)}{1+\gamma^2_t(x)}\tilde{N}({\rm d}x, {\rm d}t) \bigg].
    \label{symmetric proportional quotient rule}
    \end{align}
    \end{Lemma}
    \vspace{0.2cm}
    \noindent The corresponding results for $n$-dimensional L\'evy-Ito process are straightforward. 
    
    \begin{acknowledgements}
    \noindent The authors wish to thank D.~C.~Brody, A.~Ciatti, S. Jaimungal and L.~S\'anchez-Betancourt for helpful discussions. We are also grateful for the helpful comments of an anonymous referee. GB acknowledges support from Timelineapp Tech Ltd, Basildon.
    \\ \\
    \end{acknowledgements}

    \noindent {\bf References}\\
    
    \end{document}